\documentclass[conference]{IEEEtran}
\IEEEoverridecommandlockouts
\usepackage{cite}
\usepackage{amsmath,amssymb,amsfonts}
\usepackage{algorithmic}
\usepackage{graphicx}
\usepackage{textcomp}
\usepackage{xcolor}

\usepackage{amsthm}
\theoremstyle{definition}
\newtheorem{definition}{Definition}[section]

\usepackage{mathtools}
\DeclarePairedDelimiter{\ceil}{\lceil}{\rceil}

\usepackage{url}
\usepackage{breakurl}
\usepackage[breaklinks]{}

\usepackage{cleveref}
\crefformat{section}{\S#2#1#3} 
\crefformat{subsection}{\S#2#1#3}
\crefformat{subsubsection}{\S#2#1#3}
\Crefname{figure}{Fig.}{Figs.}

\usepackage[linesnumbered,ruled,vlined]{algorithm2e}

\SetCommentSty{mycommfont}
\SetKwInput{KwInput}{Input}                
\SetKwInput{KwOutput}{Output}              

\def\BibTeX{{\rm B\kern-.05em{\sc i\kern-.025em b}\kern-.08em
    T\kern-.1667em\lower.7ex\hbox{E}\kern-.125emX}}

\newcommand{\LL}{\textsc{L4L}}

\begin{document}

\title{Decentralizing Permissioned Blockchain with Delay Towers\\
}

\author{\IEEEauthorblockN{Shashank Motepalli,
Hans-Arno Jacobsen}
\IEEEauthorblockA{Department of Electrical and Computer Engineering,
University of Toronto\\
shashank.motepalli@mail.utoronto.ca,
jacobsen@eecg.toronto.edu}}

\maketitle

\begin{abstract}
Growing excitement around permissionless blockchains is uncovering its
latent scalability concerns. Permissioned blockchains
offer high transactional throughput and low latencies while
compromising decentralization. In the quest for a decentralized,
scalable blockchain fabric, i.e., to offer the scalability of
permissioned blockchain in a permissionless setting, we present \LL\
to encourage decentralization over the permissioned Libra network
without compromising its sustainability. \LL\ employs delay towers, --
puzzle towers that leverage verifiable delay functions -- for
establishing identity in a permissionless setting. Delay towers cannot
be parallelized due to their sequential execution, making them an
eco-friendly alternative. We also discuss methodologies to replace
validators participating in consensus to promote compliant
behavior. Our evaluations found that the cost of enabling decentralization
over permissioned networks is almost negligible. Furthermore, delay towers
offer an alternative to existing permissionless consensus mechanisms
without requiring airdrops or pre-sale of tokens.
\end{abstract}

\begin{IEEEkeywords}
Blockchain, centralization/decentralization, delay towers, verifiable
delay function, system design and analysis
\end{IEEEkeywords}

\section{Introduction}
Blockchains constitute distributed ledgers, that overcome the
challenges of centralization, for the exchange of digital assets
without trusted intermediaries in decentralized and a priori trustless
environments~\cite{nakamoto2019bitcoin}. However, the promise of
decentralization poses the challenge of needing to reach consensus on
a single immutable source of truth among untrusted parties, and not
everyone might abide by the rules of the protocol. Without loss of
generality, it is fair to assume a small percentage of (un)intentional
malicious parties that act in ways that hinder the integrity and
security of the blockchain network. Consequently, the consensus
protocols backing blockchains must tolerate Byzantine failures in
addition to benign failures. More specifically, blockchains are also
prone to Sybil attacks wherein a malicious party subverts the
consensus protocol by creating a large number of (pseudo)anonymous
identities~\cite{douceur2002sybil}. During a Sybil attack, the
malicious node can prevent creating new blocks or manipulate the
ordering of transactions. To enable Sybil resistance and ensure
security, blockchain protocols must establish and persist identities.
 
\textbf{Classification of Blockchains.} Based on how blo\-ck\-chains
establish identity and build Sybil-resistance, we classify
blo\-ck\-chains into two broad categories -- permissioned and
permissionless. Permissioned blockchains such as Hyperledger
Fabric~\cite{androulaki2018hyperledger} and Libra\footnote{On December
  1st, 2020, the Libra blockchain was renamed Diem
  blockchain.}~\cite{baudet2019state} rely on a membership service
provider (MSP), a trusted entity, to certify and authorize nodes to
access the network. This MSP moderates the network and can revoke
authorized access for malicious nodes involved in either perceived
Byzantine behavior or Sybil attacks. On the other hand, permissionless
blockchains, also often referred to as public blockchains, are more
decentralized, i.e., they do not have a centralized authority to
manage access to the network. To establish identity, permissionless
blockchains such as Bitcoin~\cite{nakamoto2019bitcoin} and
Ethereum~\cite{wood2014ethereum} leverage proof of work (PoW), wherein
the nodes compute hash puzzles. The computational power required for
PoW serves as a proxy for identity to enable Sybil resistance.
%
%
%
%
The nodes with high computational resources would have a higher
probability of proposing the next block to be appended on the
decentralized ledger. The nodes in the network reach a consensus on
the proposed block if there are no conflicts. In case of conflict, the
longest chain is chosen as the valid
ledger~\cite{nakamoto2019bitcoin}. It is worth noting that block
proposal time is different from time to reach finality
(confirmation). For instance, in Bitcoin, a probabilistic agreement is
used to reach finality; it takes approximately an hour to reach finality
after a block proposal~\cite{garay2015bitcoin,
  nakamoto2019bitcoin}. Unlike permissionless blockchains with PoW,
permissioned blockchains tend to use Byzantine fault-tolerant (BFT)
protocols to reach consensus~\cite{baudet2019state,
  androulaki2018hyperledger}. BFT protocols have deterministic
finality and tend to reach finality faster, i.e., when a quorum of
nodes agrees upon a proposed block, and they offer better scalability
than PoW~\cite{yin2019hotstuff}. We consider a system to be more
scalable if it offers higher throughput (transactions/second) and
faster finality (transaction confirmation time).
%
%
For comparison, Bitcoin and Hyperledger have a throughput of around
seven and a few thousand transactions per second,
respectively~\cite{sousa2018byzantine}. An overhead exists to build
Sybil resistance and enable decentralization, leading to a trade-off
between decentralization and scalability of
blockchains~\cite{zhang2018towards, bez2019scalability,
  xie2019survey,chauhan2018blockchain}. An increase in
decentralization often resulted in lower scalability, i.e., low
throughput and high finality time, see Table.~\ref{tab:DLTecosystem}.

\begin{table}[]
\caption{Comparison of distributed ledger technologies}
\label{tab:DLTecosystem}
\centering
\begin{tabular}{|l|l|l|}
\hline
                          & \textbf{\begin{tabular}[c]{@{}l@{}} Hyperledger or \\ Libra\end{tabular}} & \textbf{\begin{tabular}[c]{@{}l@{}}Bitcoin or\\  Ethereum\end{tabular}} \\ \hline \hline
\textbf{Type}                                                                 & Permissioned                                                                        & Permissionless                                                                    \\ \hline
\textbf{Consensus}                                                                 & BFT                                                                        & PoW                                                                    \\ \hline

\textbf{Throughput}                                                                & High                                                                         & Low                                                                    \\ \hline
\textbf{Finality}                                                                     & Fast     & Slow                                                                   \\ \hline
\textbf{Decentralization}                                                                  & Low                                                                          & High                                                                   \\ \hline
\end{tabular}

\end{table}

\textbf{The Quest.} The problem this paper addresses is characterizing
a blockchain that encourages decentralization without compromising
scalability; in other words, achieving consensus without trusted
intermediaries at scale. This problem is considered in the literature
as the quest for the scalable blockchain fabric (e.g.,
~\cite{vukolic2015quest}). This quest for decentralized, scalable
consensus is uplifting because successful expeditions could take us
one step closer to the massive adoption of blockchains.

\textbf{Expeditions.} There are many ongoing expeditions for this
quest (e.g., ~\cite{zhou2020solutions, chauhan2018blockchain, kim2018survey}). Broadly, we can achieve
decentralized scalable consensus by either scaling permissionless
blockchains or decentralizing permissioned blockchains. Lately,
multiple on-chain and off-chain solutions to scale permissionless
blockchains are being developed, such as
sharding~\cite{kokoris2018omniledger, zamani2018rapidchain},
zero-knowledge and optimistic rollups~\cite{biel_etal21, buterin_2019,
  kalodner2018arbitrum}, side chains~\cite{poon2017plasma},
bidirectional payment channels~\cite{poon2016bitcoin}, and directed
acyclic graph-based ledgers~\cite{li2020decentralized,
  alexander2018iota}.

Here, we explore the latter approach of decentralizing permissioned
blockchains. We choose to extend BFT protocol-based blockchains to
decentralized settings because they better address scalability
challenges, and we only need to introduce decentralization. Though
this sounds straightforward, making a BFT protocol permissionless
raises a plethora of challenges. Firstly, a malicious party could join
the network without a membership service provider and create many
anonymous identities to execute a Sybil attack. Secondly, BFT
protocols need a defined set of nodes, and they have to be defined
without a trusted intermediary. Having failed nodes would increase the
time to reach consensus; for instance, if a crashed node is chosen to
propose a block, it results in a timeout instead of a new
block~\cite{zhang2021prosecutor}. Also, BFT protocols do not
scale-out, i.e., with the increase in the number of nodes, lower
throughput, and higher latency were
recorded~\cite{yin2019hotstuff}. Thirdly, incentive mechanisms are
needed to reward honest behavior and punish bad behavior for not
contributing to the protocol~\cite{zhang2021prosecutor,
  motepalli2021reward}.

\textbf{\LL\ Protocol.} In this paper, we describe \LL\ to
decentralize a permissioned blockchain that uses a BFT protocol for
consensus. Specifically, we enable decentralization over the
permissioned Libra blockchain~\cite{libra2019}. We believe that a
protocol should not have a centralized MSP to be decentralized. To
achieve decentralization without MSP, we must establish identities in
a permissionless setting. To build persistent identities without heavy
PoW computations and pre-sale of native tokens, we introduce
\textit{delay towers}. Delay towers are puzzle towers~\cite{dominic}
that act as a \textit{proof of elapsed time} (PoET) leveraging
verifiable delay functions (VDF)~\cite{boneh2018verifiable}. A VDF is
a cryptographic delay function that is more eco-friendly than PoW
because it cannot be parallelized, with no substantial benefit in
using more computational resources.
%
\LL\ uses VDFs for building delay towers.  Each node locally executes a
VDF to generate proofs at regular intervals, and these proofs are
chained to build delay towers, i.e., each proof executes from the hash
of the previous proof. In addition, \LL\ reconfigures nodes
participating in consensus at regular intervals by punishing unwanted
behavior, for example, eliminating failed nodes. The nodes which
participate actively, i.e., by participating in PoET and attesting
blocks to reach consensus, are given preference to validate
transactions during reconfiguration. Though this approach might have a
few similarities to proof of stake (PoS) as in
Tendermint~\cite{buchman2016tendermint} or
Algorand~\cite{gilad2017algorand} due to usage of a BFT protocol, they
are not the same because \LL\ relies on delay towers and not staking of
native assets to establish identities. Consequently, \LL\ can be a
fairer alternative to airdrop or pre-sale of native tokens for
bootstrapping blockchain networks.

\textbf{Contributions.} The key goal of \LL\ is decentralizing a
permissioned blockchain. On this journey, \LL\ contributes the following
principle ideas:

\begin{enumerate}
	\item Delay towers to encourage persistent identities in
          permissionless settings;
	\item Mechanisms to reconfigure nodes participating in
          consensus to engender compliant behavior;
	\item More eco-friendly and sustainable alternatives to
          existing consensus protocols in permissionless settings;
    \item Mechanisms to bootstrap a blockchain network without
          token sales, airdrops, or susceptibility to mining attacks.
\end{enumerate}

\textbf{Outline.} The paper is organized as follows. The next section
introduces blockchain basics, VDFs, and the Libra blockchain. Having
discussed how to construct VDFs,~\cref{L4L-section} addresses ways of
establishing identity using delay towers in \LL. This section further
describes the reconfiguration of the nodes participating in the BFT
consensus at regular intervals in \LL. We evaluate the cost of
decentralization using delay towers from the perspective of both the node and
network in~\cref{evaluations}. Finally, we conclude with future
directions in~\cref{conclusion}.

\section{Background}
In this section. we introduce some basic blockchain formalism and
introduce verifiable delay functions (VDF), the building blocks of
delay towers; both of which are used in subsequent parts of this
paper. We end this section with an overview of the permissioned Libra
blockchain.

\subsection{Blockchain Notation}

The formalism we introduce here, serves us to characterize the \LL\
blockchain that is based on a BFT consensus protocol; the same
formalism may not readily apply to other blockchain models. A
blockchain is comprised of a distributed ledger $L$ composed of blocks
$B_i$, represented by \begin{math} L_n = \{B_0, B_1, B_2, .....,
  B_n\} \end{math}. The genesis block is the first block on the ledger
and is denoted by $B_0$. Each block $B_{k+1}$, except the genesis
block $B_0$, extends the current state of the ledger $L_{k}$ by
appending the hash of the latest committed block, $hash(B_k)$. The
memory pool, denoted as $mem$, stores the list of transactions that
are not yet published onto the ledger, represented by \begin{math} mem
  = \{tx_{j} | tx_{j} \notin L\} \end{math}. The block proposers pick
a set of transactions $tx_j \in mem$ from the memory pool to create
blocks as follows \begin{math} B_i=\{tx_{0}, tx_{1}, tx_{2},
  ... tx_{q}\} \end{math}. A block $B_{k+1}$ is valid if and only if
every transaction in that block $B_{k+1}$ is valid, and the block
contains the hash of the latest committed block $B_{k}$.

Let \begin{math}U = \{u_1, u_2, ..., u_w\}\end{math} be the users of
  this blockchain network. Each user $u_i$ has a public-private key
  pair $\{pk_i, sec_{i}\}$. A message $msg$ endorsed by user $u_i$,
  using private key $sec_{i}$, is represented as $sig_i(msg)$ and one
  can use the public key $pk_i$ to check if $sig_i(msg)$ is indeed
  signed by $u_i$.
  %

Let \textit{N}$ = \{n_1, n_2,...\}$ be the set of nodes that are
interested in listening to the blocks $B_{k}$ on the network. The full
nodes $f \subseteq N$ are the subset of nodes that do listen to
blocks $B_{k}$ on the network to replicate the ledger $L_k$ locally,
and they act as auditors to the ledger fostering decentralization. To
summarize, nodes become full nodes by initializing and tracking the
evolving ledger state.
%

The process of running delay towers is called PoET mining, and the
nodes that PoET mine are called miners. Note that mining in this paper
does not refer to PoW-mining. Furthermore, mining in itself is not
adding utility to the blockchain network, as mining does not always
mean participating in the consensus protocol but means a miner is building a persistent identity in the blockchain.


The validator set consists of nodes that participate in the consensus
protocol, i.e., validate the correctness of blocks, represented
by \begin{math}\textit{VS} = \{v_1, v_2, ..., v_t\}\end{math} and \textit{VS} $\subseteq f \subseteq$ \textit{N}. For
  instance, if a sender of a transaction uses an incorrect signature
  or does not have the required minimum account balance, the block
  containing that transaction is not valid. Since BFT protocols
  generally require $3f+1$ validators where up to $f$ of them could be
  faulty, we assume that there are at least 4 validators, i.e.,
  cardinality of validator set is always greater than or equal to four
  $|\textit{VS}| \geq 4$. A validator quorum $\mathbb{Q}$ is attained
  if a block is endorsed by the ceiling of at least two-thirds of the
  validator set, \begin{math} |\mathbb{Q}| \geq
    \ceil*{(2/3)*|\textit{VS}|} \end{math}.

We define a loosely synchronized clock function $t$ based on UNIX epoch
timestamps. After each fixed interval $t'$ in $t$, when $t \bmod t'
=0$, we have a new epoch $E$, represented by a natural number. In
other words, an epoch refers to a fixed time interval. The validator
set \textit{VS} remains constant for the duration of an epoch;
however, it can be different for different epochs. The concept of
epochs helps the protocol to reset its validator set \textit{VS} by
replacing the failed validators at regular intervals. In the current trails
of \LL, the length of an epoch is set to a day (24 hours).

\subsection{Verifiable Delay Functions}
\label{ssec:vdf}
We use verifiable delay functions (VDFs) as a proof of elapsed time
(PoET) to establish persistent identities. A VDF is a cryptographic
construct that intuitively speaking slows things down. A VDF takes a
specified number of sequential steps to be evaluated. Here, we go
through the formal specification and properties of VDFs as defined by
Boneh et al.~\cite{boneh2018verifiable}. A VDF is a set of three
functions, \texttt{VDF=(setup,eval,verify)}.
\begin{enumerate}
    \item \texttt{setup}(\texttt{security,t}) $\to$ \texttt{pp}. the setup takes in
      security parameters $security$ including pre-defined puzzle
      difficulty, and time-bound \texttt{t} to output the public
      parameters \texttt{pp}. These public parameters are unique to
      each participant and they define the input space $X$ and output
      space $Y$.
    \item $\texttt{eval}(\texttt{pp,x}) \to (\texttt{y,p})$. this is a
      deterministic function that takes time-bound \texttt{t} sequential steps to
      compute ($ f\colon X\mapsto Y$). This function consumes the
      public parameters $pp$ and an input $x$ from the input space
      \texttt{x} $\in$ $X$ to output the computation result in the
      output space \texttt{y} $\in$ $Y$ and the proof of computation
      \texttt{p}. This is the delay component in the VDF.
    \item $\texttt{verify}(\texttt{pp,x,y,p}) \to \{\texttt{true},
      \texttt{false}\}$. this is a Boolean function to verify the
      correctness of the output \texttt{y} and proof \texttt{p}.
\end{enumerate}
The \texttt{eval} function of VDF adheres to two important properties
- uniqueness and sequentiality.
\begin{enumerate}
\item Uniqueness: $\forall x \in X, \exists y \in Y$:$f(\texttt{pp,
  x})$=$(\texttt{y, p})$. For every defined input \texttt{x}, the
  output of the \texttt{eval} function \texttt{y} is unique and
  deterministic. However, the proof $p$ generated during evaluation might
  vary.
\item Sequentiality: The \texttt{eval} function takes at least
  time-bound \texttt{t} steps to compute, irrespective of parallelization
  capabilities. In other words, any parallel or random algorithm
  cannot estimate the output in less than \texttt{t} steps.
\end{enumerate}

Having defined the properties of the \texttt{eval} function, we study
candidate solutions. Time-lock puzzles are slow functions that involve
computing an inherently sequential
function~\cite{rivest1996time}. Boneh et
al.~\cite{boneh2018verifiable} proposed a generalization of time-lock
puzzles as a candidate for the \texttt{eval} function. The computation
of repeated squaring in a group of unknown order would take \texttt{t}
steps, even on a machine with parallelization
capabilities~\cite{boneh2018survey}. This function is as follows:
\begin{equation}
f(x) = x^{2^t} \mod \mathbb{N}
\end{equation}
The final step in the VDF construction is to quickly verify the
correctness of output and proofs. The candidates for \texttt{verify}
were presented by Wesolowski~\cite{wesolowski2019efficient} and
Pietrzak~\cite{pietrzak2018simple}.  We choose the Pietrzak scheme for
our VDF due to its performance benefits in the verification
step~\cite{attias2020implementation}.

\LL\ realizes chaining of VDF proofs to build delay towers to form
persistent identities (see~\cref{ssec:vdfpuzzle} for details).

\subsection{Libra Blockchain} \label{ssec:libra-background}
\textit{Libra} is a permissioned blockchain that has numerous
advantages. Firstly, Libra uses LibraBFT, a leader-based BFT protocol
under partial synchrony assumptions that guarantees safety and
liveness with deterministic finality~\cite{baudet2019state}. LibraBFT
relies on a quorum $\mathbb{Q}$ for consensus and, as a result,
prefers safety over block production
(liveness)~\cite{yin2019hotstuff}. LibraBFT builds on HotStuff,
offering linear communication complexity for reaching consensus,
making the throughput depend only on the network latency~\cite{yin2019hotstuff}. In fact,
HotStuff is reported to offer better scalability (higher throughput
and lower latency)~\cite{alqahtani2021bottlenecks} than
PBFT~\cite{castro1999practical},
Tendermint~\cite{buchman2016tendermint}, and
Streamlet~\cite{chan2020streamlet}. Pipelining in HotStuff enables the
finality of a proposed block by the third block following the proposed
block.

Secondly, Libra uses the Move programming language for smart
contracts. The Move language offers security and formal verifiability
needed for smart contracts~\cite{blackshear2019move,
  patrignani2021robust}. Libra also has modules for network
synchronization, storage, and cryptographic primitives, among
others. Finally, the Libra project is open-source, hosted on GitHub
under the Apache 2.0 license. \LL\ aims to decentralize the Libra
blockchain to provide these benefits in a permissionless setting.

\section{L4L: Design and Implementation}
\label{L4L-section}

\textit{\LL} is a consensus protocol for permissionless blockchains
designed to foster decentralization, scalability, and
sustainability. In other words, \LL\ is a Sybil-resistant Byzantine
fault-tolerant state machine replication protocol developed by
decentralizing the general-purpose permissioned Libra blockchain
without compromising sustainability.

\begin{figure}[]
  \centering
  \includegraphics[width=\columnwidth]{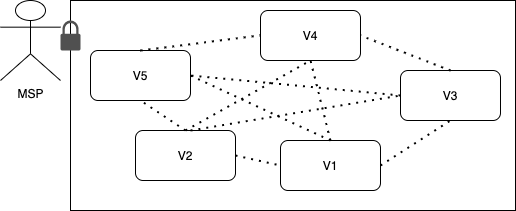}
  \caption{Membership Service Provider (MSP) decides on the validator
    set (VS) in permissioned blockchain such as Libra}
  \label{fig:MSP}
\end{figure}

A criticism of permissioned blockchains is the lack of
decentralization as access to the network is
controlled. Nonetheless, few blockchains reasonably assert that
permissioned blo\-ck\-chains are decentralized as their transaction
processing is decentralized, i.e., no single validator can decide on a
state transition of ledger $L$ by itself, and consensus requires a
quorum $\mathbb{Q}$. However, this view fails to acknowledge the
presence of a centralized authority, the MSP, who chooses the
validator set. Having an MSP administrating the network's validators
could inherit the biases of MSP in choosing validators and could
potentially result in a biased quorum $\mathbb{Q}$ assembled by an
(un)trusted intermediary. Moreover, the MSP is a point of
centralization, requiring a trusted intermediary and consequently,
causing the protocol to lose decentralization, see~\Cref{fig:MSP}.

The goal of \LL\ is to \textit{liberate the permissioned Libra blockchain} from
its decentralization-inhibiting MSP to render it permissionless. Thus,
there is a need to eliminate the requirement for a trusted
intermediary, the MSP.  However, a BFT-based network needs a fixed set
of validators to guarantee liveness and safety. An MSP plays a crucial
role in effectively administering the validators in the network, i.e.,
the MSP approves requests for joining the validator set \textit{VS}
using the identity of that node in the real world. Perhaps the most
severe disadvantage of eliminating the MSP is that there is no
identity associated with candidates in the validator set, and this
identity has to be established by other means in a permissionless
setting.

Let us review the existing practices in establishing identity
(building Sybil resistance) in permissionless blockchains. PoW
blockchains use computational power as a proxy for identity, raising
concerns about eco-friendliness~\cite{truby2018decarbonizing,
  sedlmeir2020energy}.
%
Another prominent approach is proof of stake (PoS), wherein the validators stake in
their assets as native tokens to establish identity.
%
Nevertheless, this approach requires token distribution such as
initial coin offerings (ICOs) or airdrops. There is a need for
establishing identity in a permissionless setting, and
delay towers could be a promising alternative to existing approaches.

\subsection{Delay Towers as Proof of Elapsed Time}
\label{ssec:vdfpuzzle}

To establish the persistent identities required for BFT-based networks,
\LL\ introduces the concepts of delay towers. Drawing inspiration from
Sybil-resistant network identities from dedicated hardware
in~\cite{dominic}, delay towers extend the notion of puzzle towers
with VDFs. The usage of VDFs addresses the sustainability challenges
of PoW puzzles, such as susceptibility to mining attacks or
environmental concerns.

Delay towers are a sequential series of sequential proofs. The process
of building delay towers is called \textit{PoET mining}, and every
node that is mining is called a \textit{miner}. All the miners form
the miner pool \textit{M} $\subseteq$ \textit{N}. Unlike PoW algorithms
that are parallelizable and probabilistic, PoET mining is sequential
and deterministic. Since VDFs cannot be parallelized, they have no
substantial benefit in better hardware such as
GPUs~\cite{boneh2018verifiable}. Each proof extends from the previous
proof to build the tower, creating sequential series of sequential
work. Delay towers enable persistent identities by providing
permissionless and non-forgeable identities with minimal
capital. Delay towers act as proof of elapsed time (PoET), and the
height of the delay tower denotes how long the miner has been
participating in the network, esentially showing commitment to the
network and providing a metric that can be used to rank candidates for
inclusion into the validator set \textit{VS} in \LL.

%

\begin{figure}[]
  \centering
  \includegraphics[scale = 0.5]{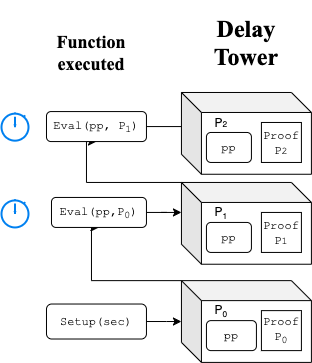}
  \caption{A delay tower is a series of VDF proofs}
  \label{fig:VDFTower}
\end{figure}
\begin{figure}[]
  \centering
  \includegraphics[scale = 0.5]{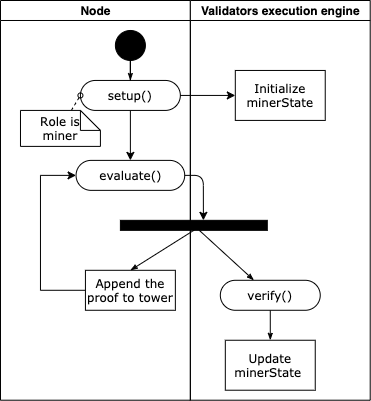}
  \caption{Activity diagram of miner}
  \label{fig:minerActivityDiagram}
\end{figure}

Every miner $m$ in \LL\ has a delay tower, represented by $T_{m} =
\{P_0, P_1, P_2, ....\}$. Each proof $P_{n+1}$, except $P_0$, builds
from its parent proof $P_n$, proving the work done by the miner after
its newest proof $P_{n}$ as shown in~\Cref{fig:VDFTower}. In terms of
implementation, \texttt{setup} and \texttt{evaluate} are executed
locally in the miner node, and the proofs are sent on-chain for
verification wherein validators execute the \texttt{verify} function,
as shown in~\Cref{fig:minerActivityDiagram}. The security parameters
\texttt{sec}, which are an input for \texttt{setup}, are fixed during
the genesis of the network.

%

A full node $f \in$ \textit{N} becomes a miner by running
\texttt{setup} and submitting its first VDF proof to initialize its
\textit{miner state} as shown in~\Cref{fig:minerActivityDiagram}. The
\texttt{setup} function receives security parameters \texttt{sec} and
public parameters \texttt{pp} as input to produce their first VDF
proof. Public parameters include the cryptographic public key $pk_f$
and the IP address of the dedicated hardware. This step links the public
key with the delay tower, making the delay towers non-transferable. The output from
\texttt{setup} is the signed parameters ($sig_f(params)$) which are
submitted to the memory pool ($mem$) from where validators check if
they constitute a valid proof to update the ledger
state then ultimately. Valid proof is defined as follows.
%
%

\begin{definition}\label{def:validProof}
\textsc{Valid Proof}. A VDF Proof $P_{k+1}$ is valid if it is either
the first proof $P_0$ with initialization parameters submitted by the
full nodes or a proof that builds on top of the hash of the latest
proof $hash(P_k)$. Moreover, a given proof returns \texttt{true} when
input to the \texttt{verify} function.
\end{definition}

If the proof $P_0$ generated from the \texttt{setup} is valid, the miner
state of that full node gets initialized, after which the full node
officially becomes a miner, joining the miner pool \textit{M}. Each
miner $m$ in the miner pool, $\{m|m \in$ \textit{M}\}, performs PoET
mining, i.e., iteratively executes \texttt{eval} locally, to submit
VDF proofs to the validator set \textit{VS}. Each proof computation,
\texttt{eval}, takes \texttt{t} steps using the hash of the previous
proof as an input to generate the output proof. To summarize, a
node executes \texttt{setup} that produces the foundation of its delay
tower and becomes a miner to execute \texttt{eval} to keep growing its
tower (see~\Cref{fig:VDFTower}).

\begin{algorithm}
\DontPrintSemicolon
  \KwOutput{true/false}
  \tcp{Authorization check}
  
  $result \gets false$
  
  \If{validSignature(addr, proof)}
    {
        \tcp{Obtain the miner state}   
        $ms \gets getMinerState(addr)$
        
        \tcp{Preliminary Checks}   
        \If{ms.hash == proof.previous\_proof\_hash and ms.height \textless  proof.height}
        {
            \tcp{Verify the VDF correctness- Pietrzak}
            \If{VDF.verify(proof)}{
                $ms.height \gets ms.height + 1$
                    
                $ms.num \gets ms.num + 1$
                
                $ms.hash \gets computeHash(proof)$
                
                $result \gets true$
            }
        }
    }
    return $result$
    
\caption{validVDFProof(addr, proof)}
\label{algo:verifyVDF}
\end{algorithm}

Thus far, we have discussed the miner $m$ view of building delay towers
$T_{m}$ via PoET mining. We now shift our focus to what happens on the
validators' end, who participate in the consensus process. When a full
node submits the output from executing \texttt{setup} locally, the
\textit{miner state} of that node is initialized as its role changes
from a full node to a miner.
%
The miner state $ms$ stores the tower height denoted by $height$, the
hash of the latest verified proof as $hash$, the number of proofs
submitted in the current epoch as $num$, $jailbit$ to signify whether
the miner is jailed, and the remaining sentence $jail\_sentence$ if it
is jailed. We discuss jailing in the following.

For all the sequential proofs $P_{k+1}$ submitted by the miner, the
validators use Algorithm~\ref{algo:verifyVDF} that runs in
O(1). The validators \textit{VS} begin by checking authorization,
i.e., the validity of the signature submitted by miner $m$. The
validators then check if the proof submitted uses the last verified
proof $hash(P_k)$ and if the tower height tallies with $height$ stored
on-chain. Finally, the validators execute the \texttt{verify} function
to verify the correctness of VDF proofs using the Pietzak scheme
(see~\cref{ssec:vdf}). If proven correct, the miner state of the
sender is updated, i.e., the validator increments the number of proofs
submitted in epoch, $num$, and the tower height, $height$. The
validator also updates the hash of the last verified proof $hash$ to
$hash(P_{k+1})$, and this process repeats for the duration of PoET
mining.
%

\subsection{Reconfiguration of Validator Set}
\label{ssec:reconfiguration}
Delay towers build a persistent identity that proves PoET, which is
non-transferable, takes time to acquire, and requires minimal
resources. However, the design described thus far fails to acknowledge
the presence of failed nodes due to crash failures or, even worse,
Byzantine behavior. If a failed validator is responsible for proposing
a new block $B_{m+1}$, the protocol would not be able to give rise to
a new block. Instead, the network at $L_{m}$ observes a timeout and
increases in latency or decreases in throughput. Even worse, if more
than one-third of validators fail, the liveness of \LL\ is compromised
due to lack of achieving a quorum $\mathbb{Q}$. Thus, the protocol
needs mechanisms to punish bad behavior for keeping the system intact.

To address these challenges, \LL\ reconfigures the validator set
\textit{VS} at regular intervals, that is, at the end of each epoch
$E$, to jail failed or non-conformat validators and to generate a new
validator set.
   
\subsubsection{Jail inactive validators} \label{ssec:jail}
As explained earlier, it is clear that failed validators do more harm
than good by affecting the network's performance, and there is a need
to punish bad behavior to maintain the integrity of the
protocol~\cite{motepalli2021reward, zhang2018towards}. Essentially,
the protocol has to answer how to measure bad behavior and punish it?
To do so, we introduce \textit{liveliness}, a metric to measure how
actively a validator node is contributing to the consensus.

\begin{definition}
\textsc{Liveliness}. The liveliness of a validator $v_i$ is the ratio
of the number of appended blocks signed by a validator to the total
number of blocks appended to the ledger $L$ in the given
epoch. Liveliness in an epoch $E$ is denoted by $l_{E}$.
\end{definition}
\begin{equation}
   l_{E}(v_i) = \dfrac{\text{blocks signed by $v_i$}}{\text{total blocks}}
\end{equation}
At the end of every epoch, \LL\ measures the liveliness of all the
validators. All the validators that do not meet the liveliness
threshold $\pi$ are considered to have failed. \LL\ punishes the failed
validators by jailing them, i.e., failed validators lose their role as
a validator for the upcoming epoch and their role is reverted back to
miners,
and they go to jail \textit{J} for jail sentence $\psi$, measured in
number of epochs $E$. To get out of jail, the jailed miners should be
PoET mining a threshold $\mu$ number of proofs every epoch for the period of the jail sentence, set as a global
constant, $\psi$. Algorithm~\ref{algo:jail} takes time O($n$) to jail
failed validators where $n$ is the cardinality of the validator set
$|\textit{VS}|$. The $jailbit$ and $jail\_sentence$ are variables of
the miner state.

%
\begin{algorithm}
    \DontPrintSemicolon
    \ForEach{$val$ in \textit{VS}}{
        $ms \gets$ $getMinerState(val)$
        
        \tcp{Check if threshold liveliness is met}  
        \If{$L_{E}(val) \leq \pi$} {
            $ms.jailbit \gets 0$
            
            $ms.jail\_sentence \gets \psi$
        }
    }
\caption{jailFailedValidators( )}
\label{algo:jail}
\end{algorithm}

\subsubsection{Select the top $\mathbb{N}$ validators} \label{ssec:topNvalidators}
Now that we jailed the failed validators, the next step is to choose
the validator set from the candidate pool called validator universe
%
\textit{VU}. $VU$ is a subset of nodes from the miner pool
(\textit{VU} $\subseteq$ \textit{M}) who state their interest in
becoming validators in the next epoch $E+1$ by mining the threshold
$\mu$ number of proofs in the current epoch $E$, stored as $num$ in
the miner state. The extraction of nodes from the validator universe
\textit{VU} involves excluding jailed miners and resetting the number
of proofs $num$ submitted in the epoch to zero. This step takes 
linear time in the cardinality of the miner pool \textit{M} as shown in
Algorithm~\ref{algo:getValidatorUniverse}.

\begin{algorithm}
    \DontPrintSemicolon
    \KwOutput{\textit{VU}}
    \KwData{\textit{M}}
    $VU \gets$ $\emptyset$
    
    \ForEach{$miner$ in \textit{M}}{
        $ms \gets$ $getMinerState(miner)$
        
        \If{$ms.num> \mu \And miner \notin$ \textit{J}} {
            $VU.insert(miner)$
        }
        
        $ms.num \gets 0$ 
    }
\caption{getValidatorUniverse( )}
\label{algo:getValidatorUniverse}
\end{algorithm}

Although all candidates in the validator universe \textit{VU} are
compliant, having everyone in the validator set is not ideal due to
the limitations of BFT protocols. BFT protocols do not scale-out,
i.e., with the increase in the cardinality of the validator set,
throughput drops, and latency increases~\cite{kokoris2018omniledger,
  yin2019hotstuff}. To counter these challenges, \LL\ caps the
cardinality of the validator set \textit{VS} to $\mathbb{N}$,
$|\textit{VS}| \leq \mathbb{N}$ and fills these slots by non-failed
validators. There is a tradeoff here: having a high $\mathbb{N}$ leads
to scalability challenges, and having a too low $\mathbb{N}$ leads to
loss of decentralization. In the current implementation, $\mathbb{N}$
is set to a hundred based on previous experimental
results~\cite{yin2019hotstuff}.

Since the candidates in the validator universe \textit{VU} might be
more than the cardinality of the validator set $\mathbb{N}$, we need a
mechanism to select candidates; here, a ranking mechanism is
used. \LL\ uses the tower height to rank the candidates due to various
advantages. Firstly, tower height is an easy to compute and
deterministic metric to measure how long the miner has been actively
PoET mining. Secondly, it is linear, providing only a minimal
advantage to genesis nodes compared to PoS mechanisms. Moreover,
building delay towers is permissionless, requiring minimal capital. A
new validator set is proposed using
Algorithm~\ref{algo:proposeValidatorSet} that takes time linear in the
order of \textit{VU}.

\begin{algorithm}
    \DontPrintSemicolon
    \KwOutput{VS}
    \KwData{\textit{M}}
    $VS \gets$ $\emptyset;$
    $i \gets $ 0
    
    $VU \gets$ getValidatorUniverse()

    sort $VU$ on $w$
    
    \tcp{Select top $\mathbb{N}$ validator candidates}  
    \While{$i \textless \mathbb{N}$ and $i \textless len(\textit{VU})$}{
        $val \gets \textit{VU}[i]$ \\
        \textit{VS}$.insert(val)$ \\ 
        $i \gets i + 1$
    }
\caption{proposeValidatorSet( )}
\label{algo:proposeValidatorSet}
\end{algorithm}

With the output from the above algorithm, \textit{\LL} reconfigures
the validator set \textit{$VS_E$} to the proposed validator set
$\textit{VS}_{E+1}$. On successful reconfiguration, the new epoch
$E+1$ begins with the updated validator set $\textit{VS}_{E+1}$
without any downtime imposed on the network.

\subsection{Lifecycle of a Node}
\label{ssec:lifecycle}
\begin{figure}[]
  \centering
  \includegraphics[width=0.5\textwidth]{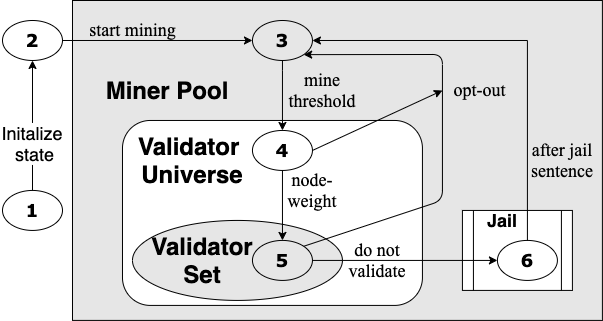}
  \caption{The lifecycle of a node}
  \label{fig:stateDiagram}
\end{figure}

\begin{table}[]
\caption{States of the nodes in the \LL\ protocol}
\label{tab:lifecycle}
\centering
\begin{tabular}{|l|l|l|l|l|}
\hline
\textbf{State} & \textbf{Role} & \textbf{Set}  & \textbf{\begin{tabular}[c]{@{}l@{}}PoET \\ Mining\end{tabular}} & \textbf{\begin{tabular}[c]{@{}l@{}}BFT \\ Validating\end{tabular}} \\ \hline \hline
1              & Node          & \textit{N}         & OFF                                                             & N/A                                                                \\ \hline
2              & Full Node     & \textit{N}         & OFF                                                             & N/A                                                                \\ \hline
3              & Miner         & \textit{M}    & ON/OFF                                                          & N/A                                                                \\ \hline
4              & Miner         & \textit{VU}    & ON                                                              & N/A                                                                \\ \hline
5              & Validator     & \textit{VS} & ON                                                              & ON                                                                 \\ \hline
6              & Miner         & \textit{J}          & ON/OFF                                                          & N/A                                                                \\ \hline
\end{tabular}
\end{table}

This section summarizes the \LL\ protocol by going through a node's
lifecycle and responsibilities as depicted in~\Cref{fig:stateDiagram}
and~\Cref{tab:lifecycle}. Full nodes $f \subseteq N$ refer to the subset of nodes
\textit{N} that initialize their ledger $L$ and listen to incoming
blocks, transitioning from State~1 to~2
in~\Cref{fig:stateDiagram}. All full nodes that execute \texttt{setup}
become miners and enter the miner pool \textit{M}, transitioning from
State~2 to~3 in~\Cref{fig:stateDiagram}. The miners who mine a
threshold $\pi$ number of proofs in an epoch \textit{E} are stating
their interest to become validators in the following epoch
\textit{E+1}. These miners are candidates for the validator set for
the following epoch \textit{$VS_{E+1}$}; they form the validator
universe \textit{VU} $\subset$ \textit{M}, represented by State~4. The
validator set (\textit{VS} $\subseteq$ \textit{VU}), represented by
State~5, is chosen by ranking the miners in the validator universe
\textit{VU}. During reconfiguration, i.e., at the end of the epoch,
the failed validators who do not meet the liveliness threshold $\pi$
are sent to jail \textit{J} as captured by State~6, and the validators
who do not meet the PoET mining threshold $\mu$ return to become
miners as captured with State~3. To get out of the jail, i.e., the
transition from State~6 to State~3, the miners have to PoET mine for
$\psi$ jail sentences.

\section{Evaluations}
\label{evaluations}

\begin{figure}[]
  \centering
  \includegraphics[width=0.45\textwidth]{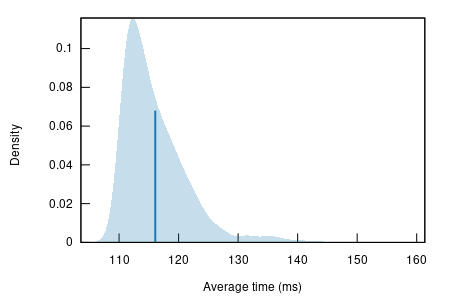}
  \caption{Probability distribution of execution times of 1500
    \texttt{verify} iterations, the line represents the mean}
  \label{fig:vdf-verify}
\end{figure}

\LL\ uses delay towers to enable permissionless blockchains. The delay
towers use Chia's Pietzak VDF implementation with 120 million
iterations and 512 bits for the length of prime numbers generated in
the proof. This profile has an estimated lower bound of thirty minutes
for generating each proof.

The miners \texttt{evaluate} proofs locally whereas the validators
\texttt{verify} the validity of submitted proofs. The time employed
for verifying the validity of proofs could instead be utilized for the
client's transaction processing;
%
it is the cost \LL\ pays for decentralization. We benchmark the
\texttt{verify} function's execution time using 1500 samples, see
probability distribution in~\Cref{fig:vdf-verify} with a mean of
115.80~ms and a median of 114.56~ms.

\begin{figure}[]
  \centering
  \includegraphics[width=0.45\textwidth]{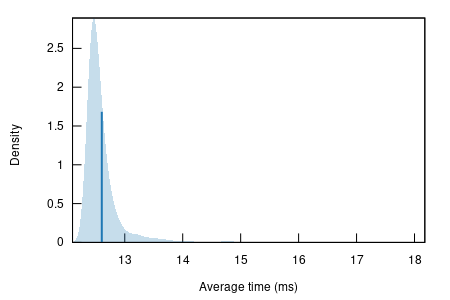}
  \caption{Probability distribution of execution times of 1500
    \texttt{verify} iterations for invalid proofs, line represents
    the mean}
  \label{fig:verify-invalid}
\end{figure}

If the length of the prime number generated as part of the proof
deviates from 512 bits, the proof is
%
rejected. Verification of invalid proofs of correct length
has a mean of 12.6~ms (11\% of valid proof computation) for 1500
samples as shown in~\Cref{fig:verify-invalid}. Assuming a validator
submits 48 valid proofs (every thirty minutes) in an epoch, the
overhead for verification is 5.52 seconds per validator per
epoch. Extrapolating this to the entire validator set
($\mathbb{N}=100$) would result in 552 seconds in an epoch
(0.0063\%). We claim that the overhead for enabling decentralization is
minimal.

\begin{figure}[]
  \centering
  \includegraphics[width=0.45\textwidth]{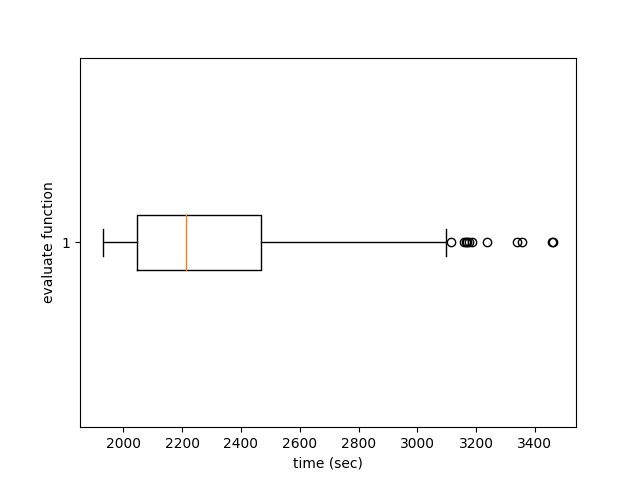}
  \caption{Box plot of 1790 samples of execution times of \texttt{evaluate} function}
  \label{fig:box-plot}
\end{figure}

We now turn to the experiments on mining the delay towers by
%
analyzing the time taken by the \texttt{evaluate} method. We collected
data from a Linux octa-core instance with 16~GB RAM (Intel DO-Regular
model). The mean and median for 1790 samples are 2362~s and 2310~s,
respectively, as shown in~\Cref{fig:box-plot}. The
25\textsuperscript{th} and 75\textsuperscript{th} percentile are at
2045~s and 2467~s, respectively. The range is wide from 1929~s to
3463~s, and subtle differences such as the compiler version could
affect the execution time. Furthermore, these insignificant benefits
in individual VDF proof computation times compound to significant tower
height over time.
%

The VDF computation plays a vital role in the security of the
network. One could impose software and hardware uniformity to address
inconsistencies, requiring every miner to use the same configurations;
however, this requirement could increase costs and be a challenge to
police without affecting the network's decentralization. Furthermore,
trusted computing is more vulnerable than PoW~\cite{chen2017security}.
%
As a countermeasure, \LL\ sets an upper limit on the tower height
growth in an epoch to mitigate any variations different hardware may
impose on VDF proof computations.  The authors recommend using the
number of epochs that a miner has mined a threshold $\mu$ of proofs,
as an alternative to pure tower height, to select the top $\mathbb{N}$
validators. This alternative would ensure all the miners who meet a
threshold to be considered equal irrespective of their CPU speeds.

\begin{figure}[]
  \centering
  \includegraphics[width=0.5\textwidth]{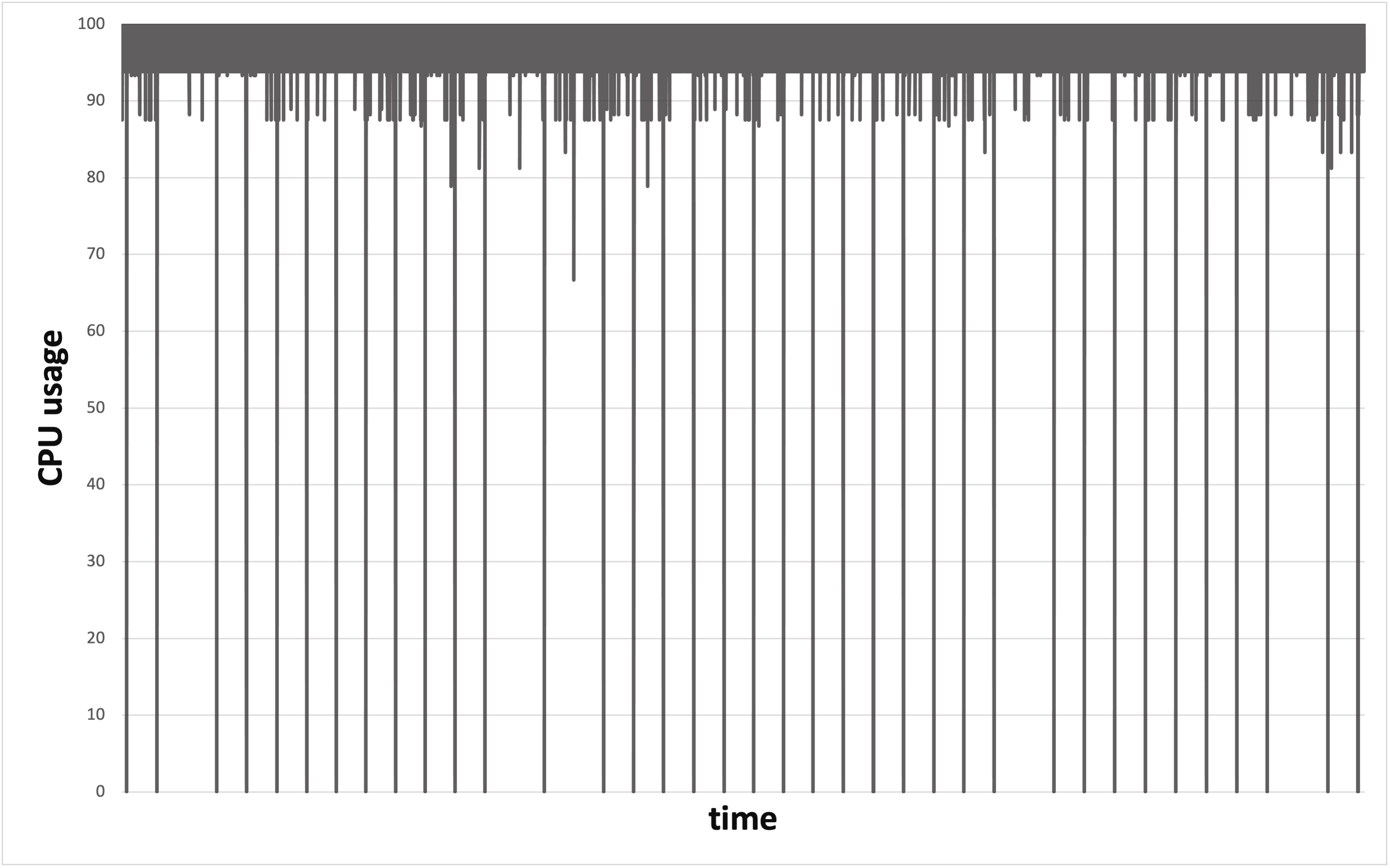}
  \caption{CPU utilization by delay tower plotted for an epoch}
  \label{fig:resources-tower}
\end{figure}

As previously stated, \LL\ uses delay towers to provide
decentralization without loss of sustainability. We measure the
computational resources of running delay towers to understand the
ecological impact. Using the same Linux instance as mentioned earlier,
we collected the CPU consumption for building a delay tower over an
epoch, as shown in the~\Cref{fig:resources-tower}. The delay tower
uses one CPU core instance in an octa-core processor, and there is a drop
in CPU utilization between the proofs. Because the process cannot be
parallelized, no more than one CPU core can ever build the same delay tower. The
delay towers use minimal system memory or network for their
processes. The data show that delay towers add minimal overhead to
both the nodes and the network.
%

\section{Conclusions}
\label{conclusion}

Intending to leverage the scalability benefits of permissioned
blockchains in permissionless settings, \LL\ decentralizes the
permissioned BFT-based Libra blockchain. To establish identities
required for BFT networks, \LL\ utilizes delay towers that are puzzle
towers with VDFs. Delay towers offer various benefits, including
lowering the barriers to entry with minimal capital requirements;
anyone with a CPU can mine delay towers. Importantly, PoET mining
delay towers are environmentally sustainable as they use minimal
computational resources, i.e., its sequential nature ensures no
advantage from parallelization. Since the growth of a delay tower is linear, 
the advantage of peers in genesis decreases over time and provides 
an opportunity to anyone interested. Furthermore, delay towers enable 
bootstrapping a network without token sales, airdrops, or being 
susceptible to mining attacks.

With the limitations of the BFT network, we showed how network
could scale from permissioned settings to permissionless setting. Let
us measure the degree of decentralization using the Nakamoto
coefficient. The Nakamoto coefficient is the minimum number of
validators needed to collude to compromise the system, i.e., affect
the safety or liveness of the network~\cite{balajis}. In a
decentralized \LL\ network of hundred validators with no Sybils, 33
validators need to collude to affect the liveness because of the
two-thirds requirement for a quorum $\mathbb{Q}$ and, hence, the Nakamoto
coefficient for \LL\ liveness is 33. For reference, the Nakamoto
coefficient for safety in Bitcoin and Ethereum is less than
five~\cite{lin2021measuring}. Furthermore, it would be interesting to
draw ideas from PoS approaches to design delegation and tower stacking
as a mechanism to scale participation in consensus beyond the limits
of BFT.

%

\section*{Acknowledgments}
The authors are indebted to 00-de-lally and Zaki Manian who
  were instrumental in making this work possible. Finally, this work
  might not be possible if it were not for the broader community of
  blockchain technology enthusiasts and practitioners behind the
  \textsc{0L.network}, an instance of the \LL\ design. The work
  presented in this paper was in part supported by NSERC.
  
\bibliographystyle{IEEEtran}
\bibliography{main-IEEE}


\end{document}